# Ninety years of X-ray spectroscopy in Italy 1896-1986

Vanda Bouché[1,2,] Antonio Bianconi[1,3,4]

[1] *Rome Int. Centre Materials Science Superstripes (RICMASS), Via dei Sabelli 119A, 00185Rome, Italy*
[2] *Physics Dept., Sapienza University of Rome, P.le A. Moro 2, 00185 Rome, Italy*
[3] *Institute of Crystallography of Consiglio Nazionale delle Ricerche, IC-CNR, Monterotondo, Rome, Italy*
[4] *National Research Nuclear University MEPhI (Moscow Engineering Physics Institute), Kashirskoeshosse 31, 115409 Moscow, Russia*

Ninety years of X-ray spectroscopy research in Italy, from the X-rays discovery (1896), and the Fermi group theoretical research (1922-1938) to the Synchrotron Radiation research in Frascati from 1963 to 1986 are here summarized showing a coherent scientific evolution which has evolved into the actual multidisciplinary research on complex phases of condensed matter with synchrotron radiation

1. **The early years of X-ray research**

The physics community was very quick to develop an intense research in Italy on X rays, after the discovery on 5th January, 1896 by Röngten in Munich. Antonio Garbasso in Pisa [1], and then in Rome Alfonso Sella, with Quirino Majorana, Pietro Blaserna and Garbasso, published several papers on Nuovo Cimento [2], starting the Italian experimental school on X-rays research in XIX century [3]. The focus was on the mechanism of the emission by the X-ray tubes, the nature of X-rays (wave or particles), the propagation characteristics in the matter, the absorption and diffusion.
Cambridge and Edinburgh at the beginning of the XX century became the world hot spots of X-ray spectroscopy with Charles Barkla who demonstrated the electromagnetic wave nature of X-ray propagation in the vacuum, similar with light and Henry Moseley, who studied the X-ray emission spectra, recorded the X-ray emission lines of most of the elements of the Mendeleev atomic table.
The wave nature of X-rays was shown in Munich in 1912 by Max Von Laue who discovered the X-ray diffraction in crystals. William Laurence and William Henry Bragg demonstrated the use of the X-ray diffraction to extract the predicted crystalline space groups of spatial arrangements of atoms in crystals. Moseley results provided evidence for a dramatic difference of X-ray sources from the thermal light sources since the X-ray emission spectra show only a very weak continuum spectrum and very narrow strong emission lines with wavelengths as short as 0.01 nm. The experimental results of Moseley led Niels Bohr of Copenhagen, traveling to Edinburgh, to the formulation of the atomic model for all elements of the Mendeleev table which was elaborated by Sommerfeld in Munich in 1915-1916, proposing the old Semi classical Quantum Physics.

Orso M. Corbino and Cesare Trabacchi in Rome directed their studies to the X-rays sources and to the X-ray crystallography after 1910. Novel high energy X-ray tubes for radio diagnostic and radiotherapy purposes were developed. Nella Mortara graduated in Rome in 1916 in Physics and became assistant professor of Corbino. She developed new X-ray techniques able to monitor the cathode electron gun, establishing laws on the relation between the cathodic current and the potential [4]. The interest on X rays and on γ ray emission from radioactive materials was dominated by the medical, diagnostic and therapy applications. Following the creation of the 'Institut du Radium' in Paris by Curie, the new institute called 'Radio Office', of the Internal Affairs Ministry was created in 1923 in the Physics Institute of Via Panisperna. In 1923 it was converted into the 'Public Health Physics Laboratory' under the Trabacchi's direction. In this Institute Nella Mortara directed her experimental studies to the natural radioactivity applied to Medical purposes.

In Florence X-ray spectroscopy was carried out by the Garbasso's students. Between them assistant professor Rita Brunetti, graduated in 1913, focused her interests on X-ray atomic spectroscopy. She proposed new experimental methods for high resolution X-ray spectra and diffraction [5].

## 2. X-ray Research and the new quantum mechanics: 1922-1938

In Pisa Enrico Fermi and Franco Rasetti graduated in 1922 with an X ray experiment. Their supervisor was prof. Pier Luigi Puccianti, chair of experimental Physics in Pisa since 1917, known for his measures of the X-ray wavelength. Fermi published two papers on X-rays: the first one is a review on the contributions of X-ray spectroscopy to foundation of the semi-classical Quantum theory and on experimental methods using Röntgen rays, citing the Brunetti's work [6]; the second one [7] is on his hand-made X-ray tube and his experimental results on X-ray imaging. These papers show that the 21 years old student of Pisa was already acquainted with the Bohr and Sommerfeld atomic models. He moved as post-doc in 1923 to work with Max Born in Gottingen Physics Institute, and in 1924 to Leiden with Paul Ehrenfest. In these two hot spots where the new Quantum Mechanics will be developed, he became friend of the major young actors of the 1925-1927 Quantum Mechanics revolution and in particular with Ralph L. de Krönig and H.A. Kramers. Kramers had worked on the interpretation of X-ray absorption spectra and on the continuous X-ray spectrum [8]. Krönig was interested on Compton effect and on the study of atoms-form factors in X-ray scattering, stimulated by Scherrer. In 1925 he was interested on the splitting of the X-ray emission lines identified by Sommerfeld and Landè, later assigned to spin orbit spitting. The idea of electron spin was presented by Krönig in 1925 when he was working as an assistant to Landé, some months before George Uhlenbeck and Samuel Goudsmit. Landé in fact had measured the presence of particular multiplets in the X ray spectra of heavy atoms which led Pauli and Krönig to suppose the existence of the electron 'spin', explained by Krönig as an electron rotation motion, and distinguished by Pauli by having only two possible values.

In the summer 1925 Krönig, Fermi and Edoardo Amaldi have spent their summer vacation in Val Gardena when Krönig was working on the elaboration of the idea of the spin and Pauli on exclusion principle. The discussions with Krönig allowed Fermi to propose "the Fermi statistics" on January 1926 and to enter among the fathers of the new Quantum Mechanics with Born, Heisenberg, Jordan, Pauli, and Schrödinger.

The X-ray absorption spectra of crystals attracted the interest of Wentzel and Koster. X-ray spectroscopy studies led to the famous Kramers-Krönig relations between absorption and dispersion [9]. Krönig developed the 'Krönig-Penney' periodic potential model for electrons in solids. Moreover he interpreted the weak modulation of the cross-section observed in X-ray absorption spectra of solids [10] as due to a *single scattering* of the photoelectron, photo-excited from a core level, by the nearest atoms in a crystal. This theory was improved by Krönig, Hartree, and Petersen for theoretical calculation of K-absorption spectra of Ge in $GeCl_4$ [11]. In 1970 the Krönig structure was renamed EXAFS (Extended X-ray Absorption Fine Structure).

In 1927 Corbino called E. Fermi on the chair of Theoretical Physics in Rome. In the same year R. Brunetti became full professor of experimental physics in Cagliari. Fermi gathered an excellent theorist group on fundamentals of the new quantum mechanics with interests on X-ray spectroscopy: Ettore Majorana, Giovanni Gentile Jr, Giulio Racah, Ugo Fano, Leo Pincherle and Piero Caldirola.

Majorana, in his first paper in collaboration with Gentile Jr. [12], focused on the spin-orbit splitting of Röngten lines in heavy atoms. By using the screening Fermi potential and the quantum electron Dirac theory in the Schrödinger equation, they deduced that the doubling is related with the near nucleus region where there is little electronic screening. In a second key paper Majorana interpreted the anomalous incomplete P triplets in Cd, Hg, Zn atomic spectra [13] and discussed the selection rules for the non-radiative decay, with a resulting interaction between discrete and continuous channels towards the same electronic configuration. This non radiative process is analogous to the Auger recombination of excited core levels, discovered by Lise Meitner and Auger, and explained by Wentzel by quantum mechanics [14]. This was called an auto-ionization process involving the overlap of three localized and one delocalized wave-functions. The scattering resonances between open and closed scattering channels due to overlap of localized and delocalized states predicted by the new Quantum Mechanics was object of the thesis of Majorana and of hot discussions between Majorana and Fermi.

Fano, graduated in mathematics at the Turin University with Enrico Persico in 1934, joined the Fermi group in the winter. Segré and Fermi asked him to give the Quantum Mechanics interpretation of the absorption line of two electron excitations degenerate with the continuum above the ionization potential in the 1s core level absorption spectrum in helium measured by Beutler. Fano [15] provided the theory for the Quantum resonance due to configuration interaction between open and closed scattering channels. The Fano theory allows to extract from the experimental measure of the anisotropic coefficient of the "Fano line-shape" of an absorption peak the strength of the resonance between a localized quasi stationary state and a continuum predicted by E. Majorana.

In 1934 L. Pincherle, graduated in Bologna on 1934, undertook in the Fermi group the studies on the Auger effect. Particularly Pincherle [16] compared the probability of an Auger electron photoemission to the probability of a radiative X-ray emission after a deep level excitation. His calculations show that Auger effect is more probable in light atoms, but less probable in heavy atoms.

Theorist in atomic and molecular spectroscopy research, Racah, Fano's cousin, in Florence was in close contact with Fermi group, and finally Caldirola joined the Fermi group in 1938 and was very influenced by Fano. In 1938 Enrico Fermi, Giulio Racah, Nella Mortara, Ugo Fano, Leo Pincherle, were forced to leave Italy; Ettore Majorana in 1938 and Giovanni Gentile Junior in 1942 passed away. The community of X-ray spectroscopy researchers disappeared.

### 3. X-ray research with Frascati synchrotron radiation: 1958 -1986

Twenty years later, in 1958, U. Fano proposed to Amaldi the revival of "X-ray spectroscopy" in Italy by using as X-ray synchrotron radiation source the Frascati 1 GeV electron-synchrotron but for Amaldi it was impossible to find experts in this field in Italy. Fano published in 1961 a revised updated version [17] of his Nuovo Cimento 1935 paper [15] and Madden and Codling [18] verified the prediction of the Fano lineshape by measuring high resolution core level K-edge absorption spectra of rare gases. Finally an international collaboration promoted by Fano, called "Sanità Luce", started between Yvette Cauchois at *Institut du Radium* in Paris and Mario Ageno of *Istituto di Sanità in Rome* [19,20] focused on the x-ray spectra of heavy elements.

In 1971 the group called "Solidi Roma built two new beam lines [21] The research was funded by a joint project of three institutions: LNF (Laboratori Nzionali di Frascati) of CNEN, represented by its director, GNSM (Gruppo Nazionale Struttura della Materia) of CNR represented by G. Chiarotti; and ISS (Istituto Superiore di Sanità) represented M. Ageno, with U. Fano of Chicago University as the foreign supervisor, visiting Frascati LNF in the summer semester. While scientists in USA focused in the "Krönig structure", a weak modulation of the cross-section over a large energy range, renamed "EXAFS" in 1970, the Frascati X-ray spectroscopy group formed by A. Bianconi, R. Habel (LNF) A. Balzarotti, E. Burattini, M. Piacentini, (GNSM) and M. Grandolfo (ISS) focused on the X-ray absorption near edge spectra proposing that the strong absorption peaks in the first 50 eV beyond the edge in Al $L_3$–edge absorption of polymorphic $Al_2O_3$ [22,23,24] and in Ca K-edge in biomolecules [25] are determined by *multiple scattering resonances* called also "shape resonances" i.e. quasi stationary states degenerate with continuum, as the broad peaks beyond the ionization potential in the K-edge absorption spectra of $N_2$ and $N_2O$ molecules predicted by Fano and his students Dehmer and Dill [26].

In 1974 Balzarotti, Bianconi, Burattini and Piacentini proposed the project for a facility called PULS (Progetto Utilizzazione Luce di Sincrotrone) using the 1.5 GeV Adone storage ring to prof. Amaldi and presented the project to CNR. The PULS construction spanned over five years. When in 1979 the research activity started at PULS, a new acronym XANES was coined to indicate the fact that XANES probe the selected local structure in complex materials via multiple scattering processes of the photoelectron [27,28,29,30] opening a new roadmap to investigate the active site geometry in proteins [31,32]. X-ray spectroscopy was shown to be a unique probe of correlated electronic systems [33,34,35], unconventional superconductors [36] and small atomic clusters [37]. Finally the X-ray spectroscopy in Italy reached a worldwide leader position in the field of XANES spectroscopy when the first EXAFS conference in 1982 [38] and the first Biophysics and Synchrotron Radiation conference in 1986 [39] were organized in Frascati.